# $C_BV_B$-nH complexes as prevalent defects in metal-organic vapor-phase epitaxy-grown hexagonal boron nitride

Marek Maciaszek[1,2,a], Bartłomiej Baur[1]

1 – Faculty of Physics, Warsaw University of Technology, Koszykowa 75, 00-662 Warsaw, Poland

2 – Center for Physical Sciences and Technology (FTMC), Vilnius LT-10257, Lithuania

a – Electronic mail: marek.maciaszek@pw.edu.pl

**Abstract:**

Optically active defects in hexagonal boron nitride (hBN) are promising candidates for active components in emerging quantum technologies, such as single-photon emitters and spin centers. However, further progress in hBN-based quantum technologies requires a deeper understanding of the physics and chemistry of hBN defects. In this work, we employ ab initio calculations to investigate the thermodynamic stability and optical properties of defect complexes involving carbon, boron vacancies, and hydrogen. We demonstrate that the formation of $C_BV_B$-nH complexes ($0 \leq n \leq 3$) is energetically favorable under nitrogen-rich conditions in the presence of carbon and hydrogen. The low formation energies and high binding energies of these complexes arise from the strong electrostatic attraction between the positively charged carbon substitutional defect ($C_B$) and the negatively charged hydrogen-passivated boron vacancies ($V_B$-nH). These complexes are particularly likely to form in metal-organic vapor-phase epitaxy (MOVPE)-grown samples, where growth occurs in the presence of carbon and hydrogen and is accompanied by a high density of boron vacancies. The optical properties of $C_BV_B$-nH complexes are analyzed and compared to recent photoluminescence measurements on MOVPE-grown hBN samples. In particular, we investigate the origin of the emission peaks at 1.90 eV and 2.24 eV and demonstrate that both the energies and lineshapes are consistent with hole capture by negatively charged $C_BV_B$ and $C_BV_B$-H complexes.

## Introduction

Hexagonal boron nitride (hBN) is a promising platform for emerging quantum technologies, serving as a host for point defects acting as single-photon emitters or spin centers [1,2].

However, further development of hBN-based technologies requires a deeper understanding of both "quantum" and "non-quantum" point defects that influence the material's overall properties. While extensive research has focused on specific defects in hBN, most studies have concentrated on their optical and spin properties in efforts to identify and characterize quantum emitters [3-5]. Identifying the most abundant defects under different growth conditions and understanding their characteristics are essential for optimizing hBN growth processes and tailoring the material for specific applications.

Numerous defect types in hBN have already been investigated both theoretically and experimentally, including native vacancies [6,7], antisites [8], carbon substitutions [9-11], and larger complexes involving carbon and oxygen [12-18]. However, hydrogen-related defects remain relatively underexplored. Hydrogen is known to readily interact with native defects in nitrides, passivating dangling bonds in vacancies and diffusing easily due to its low migration barriers [19,20]. It also influences the behavior of dopants and impurities. A well-known example of hydrogen's significance is its interaction with magnesium acceptors in GaN, which determines the achievable p-type doping levels in the material [21]. Theoretical studies have suggested that passivation of boron vacancies in hBN by hydrogen is strongly exothermic [22], and the interaction between hydrogen and boron vacancies has been proposed as a mechanism controlling the intensity of photoluminescence (PL) in the visible range in metal-organic vapor-phase epitaxy (MOVPE)-grown hBN samples [23]. Moreover, it has been demonstrated that hydrogen interacts much more strongly with boron vacancies than with nitrogen vacancies [22, 24], making the formation of $V_N$-nH much less likely.

Hydrogen-related defects are expected to play a significant role in hBN samples grown in the presence of hydrogen-containing precursors, as is common in MOVPE. Among various growth methods, MOVPE stands out by enabling wafer-scale, precisely controlled growth of high-quality hBN, combining the advantages of uniformity, defect engineering, and compatibility with industrial semiconductor technology [25, 26]. From the perspective of quantum technologies, an especially important feature is the ability to control the density of single-photon emitters and spin centers [27, 28]. To advance the application of MOVPE-grown hBN in quantum technologies, it is crucial to deepen our understanding of defect formation - particularly of defects whose formation is promoted by the specific conditions of MOVPE growth, such as hydrogen-related defects and their complexes with native or impurity defects.

In this study, we analyze complexes formed by hydrogen-passivated boron vacancies ($V_B$-nH) and carbon atoms substituting at boron sites ($C_B$). These complexes are of particular importance because their formation energy are especially low, leading to high concentrations at thermodynamic equilibrium. The low formation energy originates from the Coulomb attraction between negatively charged $V_B$-nH and positively charged $C_B$, both of which are abundant in hBN. This attraction also results in high binding energies, which is particularly relevant under non-equilibrium growth conditions, where complex formation is governed by kinetics rather than thermodynamics. $C_BV_B$-nH complexes are therefore expected to be prevalent in MOVPE-grown hBN, as MOVPE is a non-equilibrium process carried out in the presence of carbon, hydrogen and boron vacancies, whose presence in the samples is confirmed by positron annihilation spectroscopy (PAS). Furthermore, we address the optical peaks at 1.90 and 2.24 eV reported in MOVPE-grown hBN samples [23] and attribute them to hole capture by $C_BV_B$-H and $C_BV_B$ complexes, respectively. The calculated emission energies (2.03 and 2.24 eV) and full widths at half maximum (FWHM) of 0.55 and 0.40 eV show very good agreement with experimental data. Finally, we discuss mechanisms responsible for the increased intensity of the 1.90 and 2.24 eV peaks after annealing in a nitrogen atmosphere, proposing that a combination of dehydrogenation and boron-vacancy diffusion provides the most consistent explanation.

**Results and discussions**

a) **Simple defects – formation energies and electronic properties**

Firstly, we discuss the formation energies $H_f$ of simple defects, namely $V_B$, $V_B$-H, $V_B$-2H, $V_B$-3H, $C_B$ and $H_i$, which serve as the building blocks for the more complex $C_BV_B$-nH defects analyzed later. Fig. 1 shows the dependence of their formation energies on the Fermi level. These defects have been investigated theoretically in previous reports, e.g. [22, 29, 30], and our results are generally consistent with those studies, with minor discrepancies primarily attributable to methodological factors such as the choice of exchange-correlation functional, supercell size, or the treatment of electrostatic corrections. The formation energies of these defects depend strongly on the growth conditions and are significantly lower under N-rich conditions. Therefore, in the following discussion, we focus on results obtained under N-rich conditions unless stated otherwise. Under N-poor conditions, the formation energies are relatively high, making defect formation much less likely. For completeness, however, Figure 1 presents results for both limiting cases.

The boron vacancy ($V_B$) can exist in four charge states, ranging from -3 to 0. The charge-state transition levels are located at ε(0/-1)=1.90 eV, ε(-1/-2)=4.10 eV, and ε(-2/-3)=4.75 eV above the valence band maximum (VBM). In the -2 and -3 charge states, out-of-plane displacements of the nitrogen atoms surrounding the vacancy are observed, allowing the formation of bonds with boron atoms from adjacent layers - behavior that has also been reported in previous studies [29, 30]. For the 0, -2 and -3 charge states, low-spin configurations (singlet or doublet) are energetically favored, whereas for the singly negative charge state, the triplet state ($S = 1$) is the most stable configuration. For the neutral charge state, our calculations indicate that the doublet ($S = 1/2$) is the ground state, lying 0.41 eV below the quartet ($S = 3/2$), in agreement with [29-31] but in contrast to [22].

The formation energy of $V_B$-nH complexes decreases with increasing n. For the neutral complexes, the formation energies are 7.71 eV, 5.18 eV, 2.90 eV, and 1.00 eV for n=0 to n=3, respectively. This trend can be attributed to the formation of N-H bonds. In $V_B$-H, the hydrogen atom lies within the plane, whereas in $V_B$-2H and $V_B$-3H, two hydrogen atoms occupy out-of-plane positions. These geometric differences, resulting in steric and strain effects, account for the slightly nonuniform energy reductions between successive hydrogenations, i.e., between $V_B$-nH and $V_B$-(n-1)H. $V_B$-H is stable in the 0, -1 and -2 charge states, with charge-state transition levels at ε(0/-1)=2.12 and ε(-1/-2)=4.17 eV above the VBM. $V_B$-2H exists in the 0 and -1 charge states, with a transition level at ε(0/-1)=2.21 eV. $V_B$-3H remains neutral for all Fermi level positions within the band gap. For n≥1, $V_B$-nH complexes stabilize in low-spin configurations (singlet or doublet) for all charge states. The neutral $V_B$-H defect was previously reported to have a triplet ground state [22]; however, our calculations indicate that the singlet is energetically favored, lying 0.08 eV below the triplet.

The $C_B$ defect is a donor, with a (+1/0) charge-state transition level at 3.85 eV above the VBM. In the presence of carbon, $C_B$ is considered the most abundant donor, thereby determining the position of the Fermi level [14]. The negative charge state of $C_B$ can be stabilized when the Fermi level is very high, with ε(0/-1)=5.67 eV above the VBM. The $H_i$ is amphoteric, exhibiting a relatively low formation energy and a charge-state transition level at ε(+1/-1)=3.35 eV above the VBM. In contrast to the defects discussed above, $H_i$ has low migration barriers and is expected to be mobile even at room temperature [22]. Consequently, $H_i$ is likely to passivate vacancies or to annihilate at surfaces and step edges, and therefore, despite its low formation energy, its concentration in hBN in the form of isolated defects is expected to be very low.

### b) $C_BV_B$-nH – formation energies, binding energies, spin states

We now turn to the $C_BV_B$-nH complexes, whose formation energies are presented in Fig. 2.

The $C_BV_B$ defect has been identified as the carbon-vacancy complex with the lowest formation energy among all carbon-vacancy configurations without hydrogen [14]. Nevertheless, its formation energy in the neutral charge state remains relatively high, at 5.59 eV. Under conditions deviating from the N-rich limit considered here, this value would increase even further. The $C_BV_B$ complex can exist in three charge states (0, -1, and -2), with charge-state transition levels at $\varepsilon(0/-1)=2.96$ and $\varepsilon(-1/-2)=4.23$ eV above the VBM.

For $C_BV_B$-H and $C_BV_B$-2H, multiple orientation of $C_B$ relative to $V_B$-nH are possible. Our calculations indicate that the lowest-energy configuration corresponds to $C_B$ positioned near the dangling bond and away from the hydrogen atom. The most stable configurations are shown in Fig. 3.

Passivation of $V_B$ dangling bonds lowers the formation energies of the $C_BV_B$-nH complexes, consistent with the trend observed earlier for $V_B$-nH defects. For n=1 and n=2, the formation energies of the neutral charge states are 2.96 eV and 0.77 eV, respectively, with $\varepsilon(0/-1)$ levels located at 2.90 and 5.52 eV above the VBM.

The particularly low formation energies of the $C_BV_B$-nH complexes for n=1 and 2 can be rationalized by the Coulomb attraction between the positively charged $C_B^+$ donor and the negatively charged $(V_B - nH)^-$ acceptor. Because the two constituents occupy second-nearest rather than first-nearest neighbor sites, their interaction is predominantly electrostatic. Comparing the formation energy of $(C_BV_B - 2H)^0$ with the sum of $H_f(C_B^+)$ and $H_f[(V_B - 2H)^-]$ yields a large Coulomb stabilization energy of 2.06 eV, which corresponds to the complex binding energy. For $C_BV_B$-H and $C_BV_B$, the corresponding binding energies are 2.04 eV and 1.72 eV. A similar argument was presented in [14] to explain the low formation energy of the $C_NO_N$ complex, although in that case the binding energy is only 1.47 eV. The difference originates from the distinct spatial separation between the positive and negative charges. In both systems, the interacting donor and acceptor occupy second-nearest-neighbor sites, so their separation would be expected to be identical; however, this is not the case. In $C_NO_N$, the negative charge resides on the $C_N$, i.e., near the center of the site, whereas in $C_BV_B$- 2H it is localized on the dangling bond, which lies much closer to the donor $C_B$, effectively reducing the donor-acceptor separation.

The $C_BV_B$-3H complex does not follow the trend observed for n<3. It consists of $C_B$ and $V_B$-3H, with the latter remaining neutral for all Fermi-level positions within the band gap. Consequently, no Coulomb attraction is present, and the complex formation energy is not significantly reduced. The corresponding binding energies are 0.24, 0.07, and 0.38 eV for the +1, 0, and -1 charge states, respectively.

For all charge states of the $C_BV_B$-nH complexes, the preferred spin configurations are singlet or doublet, with one exception: in the neutral $C_BV_B$ complex, the triplet state is energetically favored. The singlet state lies only 0.01 eV higher in energy; however, calculations using the semilocal PBE functional predict a significantly larger singlet-triplet splitting of 0.28 eV.

Overall, the formation energies of the $C_BV_B$-nH and $V_B$-nH complexes under N-rich conditions are low, or even very low; consequently, under thermodynamic equilibrium, these defects are expected to be abundant.

### c) Fermi-level position

Since $V_B$-nH and $C_BV_B$-nH complexes can reach high concentrations and thereby affect the charge equilibrium in the material, it is important to discuss their influence on the position of the Fermi level.

The Fermi level is determined by the condition of overall charge neutrality: $p - n + \sum_D \sum_q q N(D,q) = 0$, where p and n are the hole and electron concentrations, respectively, and N(D,q) is the concentration of defect D in charge state q. For nonzero q, N(D,q) depends on the position of the Fermi level and can be expressed as $N(D,q) = N_{D,sites} \exp\left(\frac{-H_f^{D,q}(E_F)}{k_B T}\right)$, where $N_{D,sites}$ is the concentration of sites available for the formation of defect D. In principle, such a calculation requires accounting for all possible defects and their relevant charge states.

Usually, the Fermi level is discussed in terms of its role in determining the electron and hole concentrations. In wide-bandgap materials such as hBN, however, the free-carrier densities are often very low, and charge equilibrium is therefore primarily governed by defects. The main significance of the Fermi level thus lies in defining the equilibrium concentrations and thermodynamically stable charge states of these defects. In practice, the lack of sufficient carriers may hinder the establishment of uniform charge equilibrium, and the charge state of individual defects can instead be influenced by interactions with spatially nearby defects [32]. Nevertheless, examining the equilibrium Fermi-level position and its dependence on growth conditions remains a valuable starting point for gaining deeper insight into defect behavior and charge-state stabilization mechanisms. A more detailed discussion of the Fermi level in pure hBN, as well as in the presence of carbon or oxygen impurities, can be found in Refs. [22] and [14].

Under N-rich conditions in the presence of carbon and hydrogen, the calculated equilibrium Fermi level at T=1500 K - a typical growth temperature - lies 3.63 eV above the VBM. Consequently, the most stable charge states among the analyzed defects are $V_B^-$, $(V_B - H)^-$, $(V_B - 2H)^-$, $(V_B - 3H)^0$, $(C_B V_B)^-$, $(C_B V_B - H)^-$, $(C_B V_B - 2H)^0$, and $(C_B V_B - 3H)^+$. The dominant donor and acceptor species that primarily determine the Fermi-level position are $C_B^+$ and $(V_B - 2H)^-$, respectively. However, these are not necessarily the defects with the lowest formation energies overall; for example, neutral $V_B - 3H$ and $C_B V_B - 2H$ are energetically more favorable and are therefore expected to be more abundant.

Under N-poor conditions, the formation energies of defect complexes containing $V_B$ are relatively high. Consequently, the dominant donor and acceptor species are $C_B^+$ and $C_N^-$, respectively, and the equilibrium Fermi level is located 1.94 eV above the VBM. Notably, at this Fermi-level position, most $C_N$ defects are neutral, and only a small fraction are negatively charged, with a concentration approximately equal to that of $C_B^+$. This situation is analogous to that in hBN containing only carbon impurities, which was analyzed in detail in [14].

For the sake of precision, we note that interstitial defects such as $H_i$ were not included in our Fermi-level calculations, as their migration barriers are low and they are therefore not expected to exist in isolated form.

### d) $C_B V_B$-nH complexes in MOVPE-grown samples

In the following, we analyze the formation of $C_B V_B$-nH complexes in MOVPE-grown hBN, starting with key aspects of the growth conditions and sample characteristics determined in this work and in previous studies.

In typical MOVPE growth, boron precursors such as triethylborane (TEB, $(C_2H_5)_3B$) or trimethylborane (TMB, $(CH_3)_3B$) are used together with ammonia ($NH_3$), with growth temperatures in the range of 1100-1500 °C [25].

Important findings on boron vacancies in MOVPE-grown hBN have been obtained using positron annihilation spectroscopy (PAS) [23]. PAS is a powerful technique for probing positively charged or neutral vacancies and their complexes in solids [33]. In MOVPE-grown hBN, PAS measurements have revealed a high density of cation (boron) vacancies, particularly in samples (i) grown with a high $NH_3$/TEB flow ratio, corresponding to N-rich conditions, or (ii) exhibiting rough and disordered surface morphology, even when the $NH_3$/TEB flow ratio is low [23]. The defects most readily detected by PAS are non-passivated cation vacancies (i.e., boron vacancies) [33, 34]. Since their equilibrium concentrations predicted from calculated formation energies are relatively low, most of the observed vacancies are likely of non-equilibrium origin.

Ab initio studies of defect migration have shown that boron vacancies become mobile above approximately 1000 K [22, 35], indicating that they are mobile under typical MOVPE growth conditions.

Carbon is a common impurity in MOVPE-grown hBN. In samples grown using TEB or TMB as boron precursors, its concentration exceeds $10^{20}$ cm$^{-3}$ [25, 36], which is several orders of magnitude higher than the predicted equilibrium solubility limit [14]. This discrepancy indicates that defect formation during MOVPE growth proceeds under strongly non-equilibrium conditions.

Because MOVPE growth generally occurs under non-equilibrium conditions, defect formation is governed primarily by kinetics rather than thermodynamics, and the calculated formation energies may not directly determine the resulting defect concentrations. Nevertheless, the binding energies provide valuable insight, as they correspond to the reaction energies associated with complex formation. Importantly, both the passivation of boron vacancies by hydrogen and the association of $C_B$ with $V_B$-nH are strongly exothermic processes.

In summary, several aspects of the MOVPE growth environment strongly favor the formation of $C_BV_B$-nH complexes: (i) carbon and hydrogen are both supplied by the precursors and can be incorporated simultaneously, (ii) a high density of boron vacancies is typically present, (iii) boron vacancies remain mobile at growth temperatures, enabling diffusion of both $V_B$ and $C_B$ species, and (iv) the formation of $C_BV_B$-nH complexes from simpler defects is exothermic, as indicated by their large binding energies. Consequently, the formation of $C_BV_B$-nH complexes is highly probable in MOVPE-grown hBN.

### e) 1.90 and 2.24 eV PL peaks in MOVPE-grown hBN

We now turn to the analysis of the PL spectra of MOVPE-grown samples and their relation to the $C_BV_B$-nH complexes. We propose that the emission peaks centered at 1.90 eV and 2.24 eV, which are consistently observed in the PL spectra of hBN grown by MOVPE under various conditions, originate from hole capture by the $C_BV_B$-H and $C_BV_B$ complexes, respectively.

A broad emission band around 2 eV is commonly reported in the PL spectra of MOVPE-grown hBN [23, 27]. A detailed analysis combining Gaussian fitting with the Huang-Rhys model allows this broad feature to be decomposed into individual components [23]. Peaks at 1.90 and 2.24 eV appear systematically, irrespective of the growth conditions - such as the $NH_3$/TEB flow ratio or the growth mode (continuous flow versus flow-modulated epitaxy). Their intensity, however, is the highest for both very high $NH_3$/TEB flow ratio (100 and 133, corresponding to N-rich growth) and very low ratios (2.5), which lead to a rough surface morphology. These growth regimes also coincide with the conditions under which the largest concentrations of boron vacancies were determined by PAS. The 1.90 eV and 2.24 eV peaks often constitute the dominant contribution to the resolved emission band. The FWHM values range from 0.39 to 0.50 eV for the 1.90 eV band and from 0.32 to 0.39 eV for the 2.24 eV band. The PL spectra discussed here were recorded under excitation with a 473 nm (2.62 eV) laser.

Dąbrowska et al. suggested that the 1.90 and 2.24 eV emissions originate from donor–acceptor pair transitions, in which $C_B$ acts as the donor and $V_B$ or $V_B$-H serves as the acceptor [23]. While this interpretation captures the essential roles of carbon, hydrogen and boron vacancies, we propose a slight modification: the emissions are more consistently explained by hole capture at negatively charged $C_BV_B$ and $C_BV_B$-H complexes. The interaction between spatially separated $C_B$ and $V_B$-nH centers, as proposed in [23], appears less probable, since their orbitals are strongly localized, which would render the transition dipole moment for such a process very small.

We propose that the optical emission arises from radiative recombination following hole capture at negatively charged $C_BV_B$ and $C_BV_B$-H. Both complexes are expected to be stable in the negative charge state, as their (0/-) charge-state transition levels lie 2.96 and 2.90 eV above the VBM for $C_BV_B$ and $C_BV_B$-H, respectively, whereas under thermodynamic equilibrium and N-rich conditions the Fermi level is located 3.63 eV above the VBM. We now consider the origin of holes under excitation with a 473 nm (2.62 eV) laser. In principle, excitation at this wavelength cannot directly generate electron-hole pairs. However, carriers may still be produced through defect- or surface-assisted processes. For instance, boron vacancies illuminated at 473 nm can undergo internal excitation followed by optical hole emission. Another possible pathway involves optical hole emission from $V_B^0$. These neutral vacancies may form via photoionization of $V_B^-$ in its excited state. Assuming a zero-phonon line (ZPL) energy of 1.60 eV for the intra-defect transition in $V_B^-$ [7], the threshold for photoionization is given by $E_g - E_{ZPL} - \varepsilon(0/-1) = 2.45$ eV, while the threshold for optical hole emission from $V_B^0$ is $\varepsilon(0/-1) = 1.90$ eV. Consequently, the complete three-step optical cycle has a threshold of 2.45 eV, making its activation feasible under 2.62 eV excitation. Similar carrier-generation mechanisms involving boron vacancies in MOVPE-grown hBN have been proposed in Refs. [37] and [38].

The ZPL for optical emission via hole capture by negatively charged $C_BV_B$ is determined by its (0/-1) charge-state transition level, located 2.96 eV above the VBM (2.90 eV for $C_BV_B$-H). The emission intensity at the ZPL is expected to be very weak owing to strong electron-phonon coupling. The relaxation energy (Franck-Condon shift) $E_{rel}$ are calculated to be 0.45 eV for $C_BV_B$ and 0.60 eV for $C_BV_B$-H. Importantly, the captured hole does not originate from a delocalized valence-band state but from a bound state formed through electrostatic attraction between the negatively charged complex and the hole. Although its energy cannot be determined with high accuracy within standard DFT using computationally feasible supercells, a comparison between the ZPL calculated for neutral $C_N$ (2.47 eV [10]) and the (0/-1) transition level obtained here for the same defect (2.74 eV) suggests a hole binding energy $E_b$ of 0.27 eV. Taking this binding energy into account, the resulting vertical emission energies - corresponding to the maxima in the emission spectra - are $\varepsilon(0/-1) - E_{rel} - E_b = 2.24$ eV for $C_BV_B$, and 2.03 eV for $C_BV_B$-H, which are in very good agreement with the experimental values of 2.24 eV and 1.90 eV, respectively. The corresponding configuration–coordinate diagram for $C_BV_B$ is shown in Fig. 4a.

We now analyze the emission lineshapes of the optical transitions, which requires evaluation of the Huang-Rhys factor S, representing the average number of phonons emitted during a transition. Because the electron-phonon coupling is relatively strong for both complexes, the effective-mode approximation is appropriate [39, 40]. The calculated effective phonon frequencies $\hbar\omega_{eff}$ are 60 meV for $C_BV_B$ and 90 meV for $C_BV_B$-H, corresponding to Huang-Rhys factors $S = E_{rel}/\hbar\omega_{eff} = 7.40$ and 6.64, respectively.

The emission lineshape calculated within the single effective-mode approach [39] for $C_BV_B$ is shown in Fig. 4b and exhibits an FWHM of 0.40 eV. Owing to the strong electron-phonon coupling, the lineshape is nearly Gaussian and only weakly asymmetric. The theoretical FWHM in the limit of strong coupling at 0 K is given by $\hbar\omega_{eff}\sqrt{8\ln 2\, S}$ [40], yielding a very similar value. For $C_BV_B$-H, the FWHM of the computed spectrum is larger, 0.55 eV. Experimentally, the observed FWHM values are 0.32-0.39 eV for the 2.24 eV peak and 0.39-0.50 eV for the 1.90 eV peak, in good agreement with our calculations. It should be noted that the theoretical results correspond to 0 K, whereas the experimental spectra were recorded at room temperature. Nevertheless, since the phonon energies involved (60-90 meV) are much larger than $k_BT$ at 300 K, the influence of temperature on the linewidths is expected to be negligible.

### f) Effect of annealing on $C_BV_B$-nH complexes

In [23], additional PL measurements were performed after annealing the samples in a nitrogen atmosphere at 700 °C for t=5 min. Annealing generally enhanced the 1.90 eV and 2.24 eV peaks, most strongly in samples grown with NH₃/TEB flow ratios of 25 and 100, while the effect was much weaker for the low-ratio (2.5) sample and negligible for the flow-modulated epitaxy samples with smooth morphology. Dąbrowska et al. attributed this increase to dehydrogenation of hydrogen-passivated vacancies, such as the abundant $V_B$-3H centers [23]. The same mechanism could, in principle, also apply to the $C_BV_B$-nH complexes analyzed here. However, the hydrogen removal energies are relatively high, and such a process may therefore be inefficient at 700 °C. Below, we outline an additional mechanism that could increase the concentrations of $C_BV_B$ and $C_BV_B$-H complexes during annealing, coexisting with thermal dehydrogenation. The proposed mechanism helps to explain the difference in annealing effects between samples grown with various NH₃/TEB flow ratios or under different growth modes (continuous flow versus flow-modulated epitaxy).

We assume that during annealing, boron vacancies become sufficiently mobile to be captured by $C_B^+$ donors, forming $C_BV_B$ complexes. We consider two scenarios: diffusion of $V_B$ in the -1 or -2 charge states. The number of effective hops during annealing is estimated as $N = \Gamma_0 t \exp\left(\frac{-E_m}{k_BT}\right)$, where $E_m$ is the migration barrier and $\Gamma_0$ is the attempt frequency (typically set by the phonon frequency), which in hBN can be approximated as $10^{14}$ s$^{-1}$. The average diffusion distance L after N hops is then obtained from the root-mean-square displacement, $L = d\sqrt{N}$, where d is the hopping distance (d=2.49 Å).

Using migration barriers of $E_m$=3.09 eV for $V_B^-$ and 2.33 eV for $V_B^{2-}$ [22] yields only about three effective hops for $V_B^-$, but as many as $2.6 \times 10^4$ for $V_B^{2-}$. This corresponds to diffusion lengths of 4.3 Å for $V_B^-$ and 399 Å for $V_B^{2-}$.

However, the -2 charge state is not the energetically preferred configuration of $V_B$. Its appearance would require, for instance, electron transfer between two $V_B^-$ centers, producing $V_B^0$ and $V_B^{2-}$. Such a process is endothermic, with a reaction energy of 2.20 eV, and becomes likely only when the vacancies are in close spatial proximity.

For $V_B^-$, diffusion is strongly limited by the high migration barrier. However, comparisons of $E_m$ values for different vacancies and charge states obtained using various computational methods reveal discrepancies of up to 0.6 eV [22, 35]. If the barrier is reduced to 2.90 eV, the average number of hops of $V_B^-$ increases to 29, corresponding to a diffusion length of 13.3 Å. This length is comparable to the average separation between carbon atoms at a concentration of ~$10^{20}$ cm$^{-3}$, typical for MOVPE-grown samples. Moreover, because Coulomb attraction drives vacancy migration toward $C_B^+$, the calculated average diffusion length likely underestimates the true capture probability. Once a complex is formed, electrostatic binding suppresses further vacancy migration, resulting in neutral $C_BV_B$ complexes. Optical emission at 2.24 eV, however, requires the complex to be in a negative charge state. Under 473 nm excitation, emission of a hole into the bound state can occur, with a threshold of $\varepsilon(0/-) - E_b = 2.69$ eV, only slightly higher than the excitation energy. Another possible pathway involves the capture of an electron produced by photoionization of $V_B^-$, as discussed earlier.

The model involving boron-vacancy diffusion during annealing naturally explains why annealing predominantly affects films grown at high NH$_3$/TEB ratios: such conditions yield higher densities of both $V_B$ and $C_B$, many of which remain unpaired. In contrast, in the sample grown with a ratio of 2.5 (N-poor), the density of $C_B$ is expected to be lower, and most $C_B$ donors are already paired with the more abundant $V_B$ acceptors (i.e., before annealing). Flow-modulated epitaxial films with smooth morphology exhibit a low density of $V_B$; consequently, $C_BV_B$ and $C_BV_B$-H complexes are not abundant either before or after annealing.

Newly formed $C_BV_B$ complexes can subsequently trap hydrogen. The hydrogen may originate from partial dehydrogenation of $V_B$-3H during annealing or from grain boundaries acting as hydrogen reservoirs. Mobile interstitial hydrogen – particularly $H_i^-$, whose migration barrier is only 0.46 eV [22] – is efficiently captured by $C_BV_B$, leading to the observed increase in $C_BV_B$-H.

**Conclusions**

In this work, we have performed a comprehensive first-principles investigation of carbon-boron-vacancy-hydrogen complexes in hBN, focusing on their energetic stability and optical properties. Our results show that the formation of $C_BV_B$-nH complexes is energetically favorable under N-rich conditions, and the strong Coulomb attraction between positively charged $C_B$ and negatively charged $V_B$-nH yields low formation energies and high binding energies. Such complexes are especially likely to form during the MOVPE growth of hBN, where carbon, hydrogen, and boron vacancies coexist and remain mobile at growth temperatures.

The calculated emission energies for the $C_BV_B$ and $C_BV_B$-H defects (2.24 eV and 2.03 eV, respectively) are in good agreement with the experimentally observed photoluminescence peaks at 2.24 eV and 1.90 eV. The large Huang-Rhys factors obtained for both complexes indicate strong electron-phonon coupling, consistent with the broad emission features observed experimentally. Importantly, the calculated FWHM values (0.40 eV for $C_BV_B$ and 0.55 eV for $C_BV_B$-H) closely match the linewidths extracted from Gaussian deconvolution of the experimental spectra. Finally, analysis of annealing effects in samples grown with various $NH_3$/TEB flow ratios or under different growth modes suggests that both dehydrogenation and boron-vacancy diffusion contribute to the enhanced PL intensity observed after thermal treatment. Overall, our findings identify $C_BV_B$-nH complexes as key defects in MOVPE-grown hBN and provide a foundation for a deeper understanding of visible-range emission in this material.

**Methods**

The calculations were performed using density functional theory as implemented in the Vienna Ab initio Simulation Package (VASP) [41]. Spin polarization was included in all calculations.

The exchange-correlation energy was described using the HSE06 hybrid functional [42], with the fraction of exact exchange increased to 0.31 to reproduce the experimental band gap. The projector augmented-wave method was employed [43], and a plane-wave cutoff energy of 500 eV was used. Defect calculations were conducted using a 288-atom supercell (6x6x2). The Brillouin zone was sampled only at the Γ point. For all defect-containing supercells, atomic positions were relaxed until all forces were below 0.01 eV/Å (0.005 eV/Å for phonon-frequency calculations). To account for van der Waals interactions, the Grimme-D3 correction was applied [44].

Using this methodology, the calculated band gap was 5.95 eV. The optimized lattice constants were a=2.490 Å and c=6.558 Å, and the calculated formation enthalpy of the compound, $\Delta H_f(\text{hBN})$, was -2.89 eV. These values are in good agreement with experimental data of a=2.506 Å, c=6.603 Å, and $\Delta H_f(\text{hBN}) = -2.60$ eV [45, 46].

The formation energy of a defect D in charge state q was calculated using the expression [47, 48]:

$$H_f(D, q) = E_D(D, q) - E_H + \sum_i n_i \mu_i + q(E_V + E_F) + E_{corr}$$

where $E_D$ is the total energy of the supercell containing the defect D, and $E_H$ is the total energy of the perfect (defect-free) supercell. $\mu_i$ represents the chemical potential of element i, and $n_i$ is the number of atoms of type i added ($n_i<0$) or removed ($n_i>0$) to form the defect. $E_V$ is the energy of the valence-band maximum. $E_{corr}$ denotes the correction term accounting for electrostatic interactions between charged supercells, estimated using the Freysoldt-Neugebauer-Van de Walle scheme [49]. The charge-state transition level, ε(q/q'), corresponds to the Fermi-level position at which the formation energies of the defect in charge states q and q' are equal.

The chemical potentials define the thermodynamic growth conditions of the material. For boron and nitrogen, they must satisfy the following relations [22, 48]:

$$\mu_B + \mu_N = E(\text{hBN})$$

$$\mu_B^{solid} + \Delta H_f(\text{hBN}) < \mu_B < \mu_B^{solid}$$

$$\mu_N^{gas} + \Delta H_f(\text{hBN}) < \mu_N < \mu_N^{gas}$$

where $E(\text{hBN})$ is the total energy of hBN per formula unit ($E(\text{hBN}) = E_H/n$, with n=144 in this work).

Under N-rich conditions, $\mu_N$ reaches its maximum value, $\mu_N = \mu_N^{\text{gas}} = \frac{1}{2}E(\text{N}_2)$, while $\mu_B$ takes its minimum value, $\mu_B = \mu_B^{\text{solid}} + \Delta H_f(\text{hBN})$. For carbon, $\mu_C$ was taken as the energy per atom in diamond. For hydrogen, several reference states were considered: H$_2$, NH$_3$, and CH$_4$. Under N-rich conditions, $\mu_H$ is determined by the stability of the NH$_3$ molecule: $\mu_H = \frac{1}{3}[E(\text{NH}_3) - \mu_N]$. This corresponds to the maximum hydrogen chemical potential for which the formation of NH$_3$ or CH$_4$ remains thermodynamically unfavorable. Under N-poor conditions, $\mu_B = \mu_B^{\text{solid}}$, $\mu_N = \frac{1}{2}E(\text{N}_2) + \Delta H_f(\text{hBN})$, and the hydrogen chemical potential is constrained by the stability of the CH$_4$ molecule: $\mu_H = \frac{1}{4}[E(\text{CH}_4) - \mu_C]$.

**References**

1. Caldwell, J. D. et al. Photonics with hexagonal boron nitride. *Nat. Rev. Mater.* **4**, 552−567 (2019).
2. Aharonovich, I., Tetienne, J.-P. & Toth, M. Quantum Emitters in Hexagonal Boron Nitride. *Nano Lett.* **22**, 9227−9235 (2022).
3. Çakan, A. et al. Quantum Optics Applications of Hexagonal Boron Nitride Defects. *Adv. Opt. Mater.* **13**, 2402508 (2025).
4. Mai, T. N. A. et al. Quantum Emitters in Hexagonal Boron Nitride: Principles, Engineering and Applications. *Adv. Funct. Mater.* 10.1002/adfm.202500714 (2025).
5. Carbone, A. et al. Creation and microscopic origins of single-photon emitters in transition-metal dichalcogenides and hexagonal boron nitride. *Appl. Phys. Rev.* **12**, 031333 (2025).
6. Ivády, V. et al. Ab initio theory of the negatively charged boron vacancy qubit in hexagonal boron nitride. *Npj Comput. Mater.* **6**, 41 (2020).
7. Qian, C. et al. Unveiling the Zero-Phonon Line of the Boron Vacancy Center by Cavity-Enhanced Emission. *Nano Lett.* **22**, 5137-5142 (2022).
8. Li, S., Li. P. & Gali, A. Native antisite defects in h-BN. *Appl. Phys. Lett.* **126**, 062104 (2025).
9. Jara, C. et al. First-Principles Identification of Single Photon Emitters Based on Carbon Clusters in Hexagonal Boron Nitride. *J. Phys. Chem. A* **125**, 1325−1335 (2021).
10. Auburger, P. & Gali, A. Towards ab initio identification of paramagnetic substitutional carbon defects in hexagonal boron nitride acting as quantum bits. *Phys. Rev. B* **104**, 075410 (2021).
11. Chacon, I., Echeverri, A., Cardenas, C. & Munoz, F. Optical properties and spin states of inter-layer carbon defect pairs in hexagonal boron nitride: a first-principles study. *Phys. Chem. Chem. Phys.* **27**, 16454-16464 (2025).

12. Li, K., Smart, T. J. & Ping, Y. Carbon trimer as a 2 eV single photon emitter candidate in hexagonal boron nitride: A first principles study. *Phys. Rev. Mater.* **6**, L042201 (2022).
13. Maciaszek, M. & Razinkovas, L. Blue Quantum Emitter in Hexagonal Boron Nitride and a Carbon Chain Tetramer: a First-Principles Study. *ACS Appl. Nano Mater.* **7**, 18979-18985 (2024).
14. Maciaszek, M., Razinkovas, L. & Alkauskas, A. Thermodynamics of carbon point defects in hexagonal boron nitride. *Phys. Rev. Mater.* **6**, 014005 (2022).
15. Li, S., Pershin, A., Thiering, G., Udvarhelyi, P. & Gali, A. Ultraviolet Quantum Emitters in Hexagonal Boron Nitride from Carbon Clusters. *J. Phys. Chem. Lett.* **13**, 3150−3157 (2022).
16. Li, S., Pershin, A. & Gali, A. Quantum emission from coupled spin pairs in hexagonal boron nitride. *Nat. Commun.* **16**, 5842 (2025).
17. Babar, R., Ganyecz, Á., Abrikosov, I. A., Barcza, G. & Ivády, V. Carbon-contaminated topological defects in hexagonal boron nitride for quantum photonics. *Npj 2D Mater. Appl.* **9**, 33 (2025).
18. Benedek, Z. et al. Symmetric carbon tetramers forming spin qubits in hexagonal boron nitride. *Npj Comput. Mater.* **9**, 187 (2023).
19. Van de Walle, C. G. & Neugebauer, J. First-principles calculations for defects and impurities: Applications to III-nitrides. *J. Appl. Phys.* **95**, 3851–3879 (2004).
20. Dreyer, C. E., Alkauskas, A., Lyons, J. L., Speck, J. S. & Van de Walle, C. G. Gallium vacancy complexes as a cause of Shockley-Read-Hall recombination in III-nitride light emitters. *App. Phys. Lett.* **108**, 141101 (2016).
21. Nakamura, S., Mukai, T., Senoh, M., & Iwasa, N. Thermal Annealing Effects on P-Type Mg-Doped GaN Films. *Jpn. J. Appl. Phys.* **31**, L139 (1992).
22. Weston, L., Wickramaratne, D., Mackoit, M., Alkauskas, A. & Van de Walle, C. G. Native point defects and impurities in hexagonal boron nitride. *Phys. Rev. B* **97**, 214104 (2018).
23. Dąbrowska, A. K. et al. Defects in layered boron nitride grown by Metal Organic Vapor Phase Epitaxy: luminescence and positron annihilation studies. *J. Lumin.* **269**, 120486 (2024).
24. McDougall, N. L., Partridge, J. G., Nicholls, R. J., Russo, S. P. & McCulloch, D. G. Influence of point defects on the near edge structure of hexagonal boron nitride. *Phys. Rev. B* **96**, 144106 (2017).
25. Moon, S. et al. Metal-organic chemical vapor deposition of hexagonal boron nitride: from high-quality growth to functional engineering. *2D Mater.* **12**, 042006 (2025).
26. Pakuła, K. et al. Fundamental mechanisms of hBN growth by MOVPE. arXiv:1906.05319 (2019).
27. Mendelson, N. et al. Identifying carbon as the source of visible single-photon emission from hexagonal boron nitride. *Nat. Mater.* **20**, 321–328 (2021).
28. Iwański, J. et al. Manipulating carbon related spin defects in boron nitride by changing the MOCVD growth temperature. *Diam. Relat. Mater.* **147**, 111291 (2024).
29. Wang, V., Liu, R.-J., He, H.-P., Yang, C.-M. & Ma, L. Hybrid functional with semi-empirical van der Waals study of native defects in hexagonal BN. *Solid State Commun.* **177**, 74–79 (2014).
30. Huang, B. & Lee, H. Defect and impurity properties of hexagonal boron nitride: A first-principles calculation. *Phys. Rev. B* **86**, 245406 (2012).

31. Attaccalite, C., Bockstedte, M., Marini, A., Rubio, A. & Wirtz, L. Coupling of excitons and defect states in boron-nitride nanostructures. *Phys. Rev. B* **83**, 144115 (2011).
32. Collins, A. T. The Fermi level in diamond. *J. Phys.: Condens. Matter* **14**, 3743 (2002).
33. Tuomisto, F. & Makkonen, I. Defect identification in semiconductors with positron annihilation: Experiment and theory. *Rev. Mod. Phys.* **85**, 1583 (2013).
34. Karjalainen, A. et al. Interplay of vacancies, hydrogen, and electrical compensation in irradiated and annealed n-type β-$Ga_2O_3$. *J. Appl. Phys.* **129**, 165702 (2021).
35. Zobelli, A., Ewels, C. P., Gloter, A. & Seifert, G. Vacancy migration in hexagonal boron nitride. *Phys. Rev. B* **75**, 094104 (2007).
36. Yamada, H., Inotsume, S., Kumagai, N., Yamada, T. & Shimizu, M. Comparative Study of Boron Precursors for Chemical Vapor-Phase Deposition-Grown Hexagonal Boron Nitride Thin Films. *Phys. Status Solidi A* **218**, 2000241 (2021).
37. Wlasny, I., Stepniewski, R., Klusek, Z., Strupinski, W. & Wysmolek, A. Laser-controlled field effect in graphene/hexagonal boron nitride heterostructures. *J. Appl. Phys.* **123**, 235103 (2018).
38. Iwański, J. et al. Temperature induced giant shift of phonon energy in epitaxial boron nitride layers. *Nanotechnology* **34**, 015202 (2023).
39. Alkauskas, A., Lyons, J. L., Steiauf, D., & Van de Walle, C. G. First-Principles Calculations of Luminescence Spectrum Line Shapes for Defects in Semiconductors: The Example of GaN and ZnO. *Phys. Rev. Lett.* **109**, 267401 (2012).
40. Alkauskas, A., McCluskey, M. D. & Van de Walle, C. G. Tutorial: Defects in semiconductors—Combining experiment and theory. *J. Appl. Phys.* **119**, 181101 (2016).
41. Kresse, G. & Furthmüller, J. Efficient iterative schemes for ab initio total-energy calculations using a plane-wave basis set. *Phys. Rev. B* **54**, 11169 (1996).
42. Heyd, J., Scuseria, G. E. & Ernzerhof, M. Hybrid functionals based on a screened Coulomb potential. *J. Chem. Phys.* **118**, 8207 (2003).
43. Blöchl, P. E. Projector augmented-wave method. *Phys. Rev. B* **50**, 17953 (1994).
44. Grimme, S. Semiempirical GGA-type density functional constructed with a long-range dispersion correction. *J. Comput. Chem.* **27**, 1787 (2006).
45. Paszkowicz, W., Pelka, J. B., Knapp, M., Szyszko, T. & Podsiadlo, S. Lattice parameters and anisotropic thermal expansion of hexagonal boron nitride in the 10−297.5 K temperature range. *Appl. Phys. A* **75**, 431−435 (2002).
46. Tomaszkiewicz, I. The Enthalpy of Formation of Hexagonal Boron Nitride. *Polish J. Chem.* **76**, 891−899 (2002).
47. Zhang, S. B. & Northrup, J. E. Chemical potential dependence of defect formation energies in GaAs: Application to Ga self-diffusion. *Phys. Rev. Lett.* **67**, 2339 (1991).
48. Freysoldt, C. et al. First-principles calculations for point defects in solids. *Rev. Mod. Phys.* **86**, 253 (2014).
49. Freysoldt, C., Neugebauer, J. & Van de Walle, C. G. Fully Ab Initio Finite-Size Corrections for Charged-Defect Supercell Calculations. *Phys. Rev. Lett.* **102**, 016402 (2009).

**Acknowledgments**

We acknowledge Prof. Andrzej Wysmołek and Jakub Iwański for valuable discussions. This research was funded by the Warsaw University of Technology within the Excellence Initiative: Research University (IDUB) programme. Computational resources were provided by the Interdisciplinary Center for Mathematical and Computational Modelling (ICM), University of Warsaw (Grant No. GB81-6), and by the High Performance Computing Center "HPC Saulėtekis" in the Faculty of Physics, Vilnius University, Lithuania.

**Data availability statement**

The data that support the findings of this study are available from the corresponding author upon reasonable request.

a)

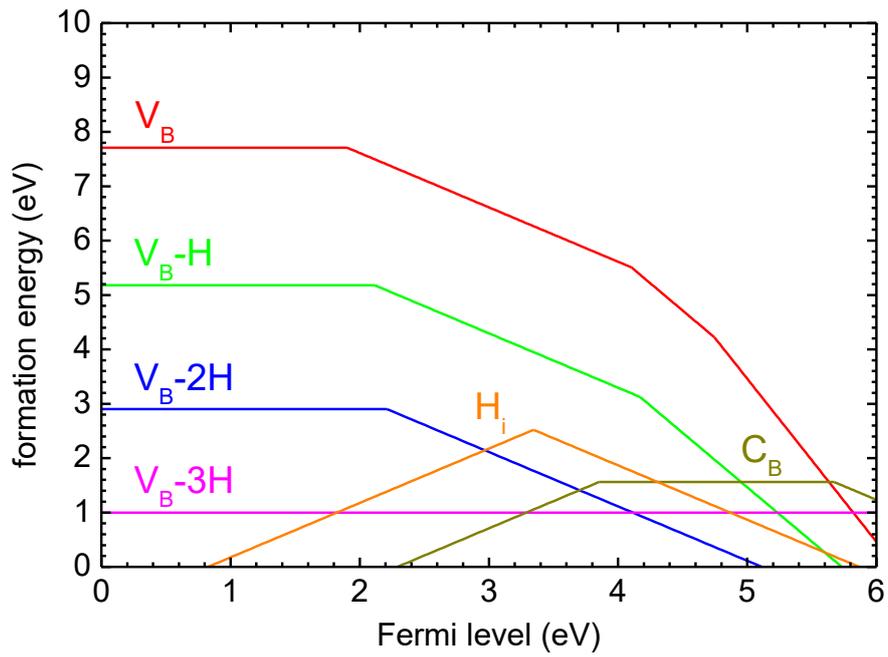

b)

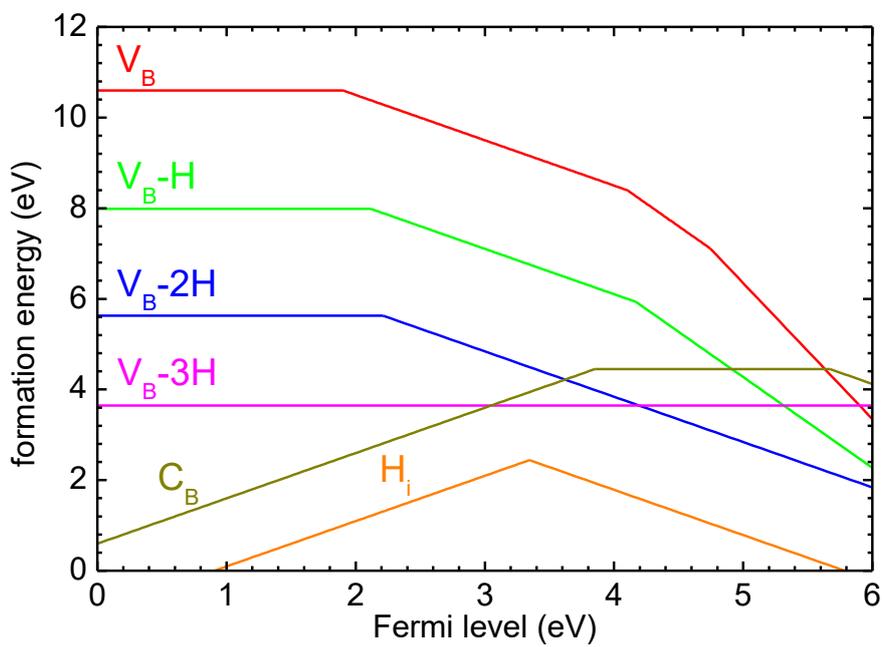

Fig. 1. Formation energies of simple defects ($V_B$, $V_B$-H, $V_B$-2H, $V_B$-3H, $C_B$, and $H_i$) as a function of the Fermi level under (a) N-rich and (b) N-poor conditions.

a)

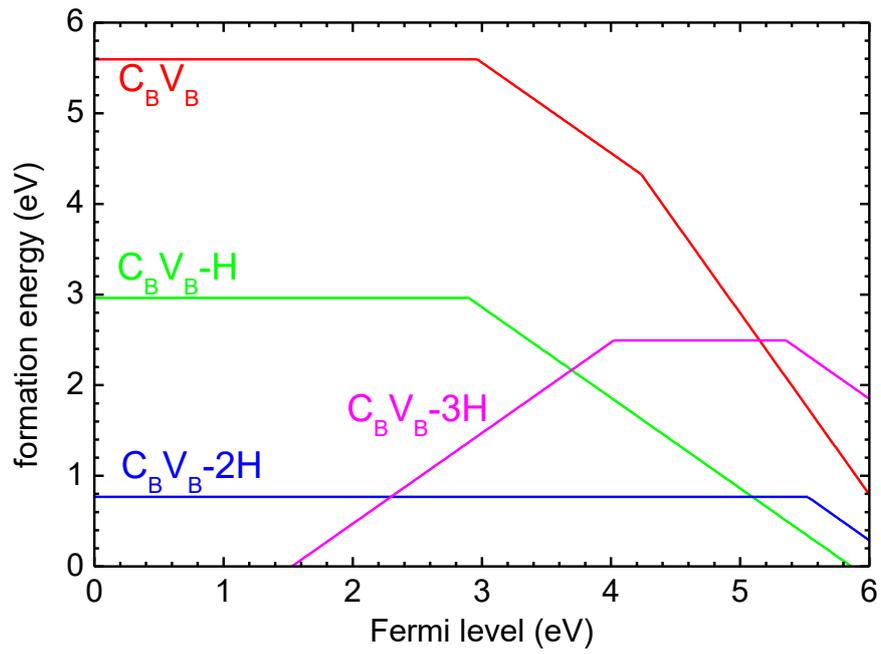

b)

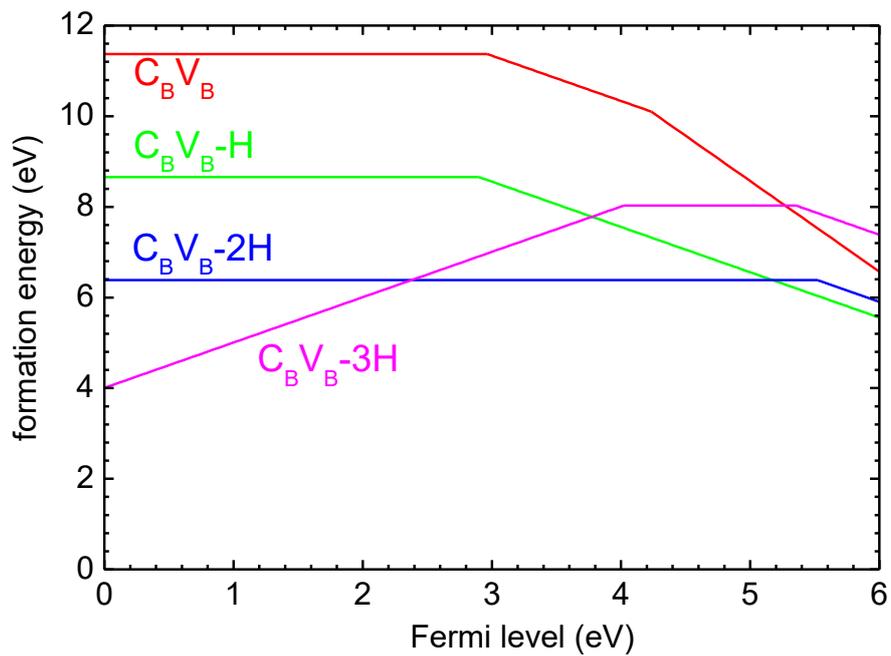

Fig. 2. Formation energies of $C_BV_B$-nH complexes as a function of the Fermi level under (a) N-rich and (b) N-poor conditions.

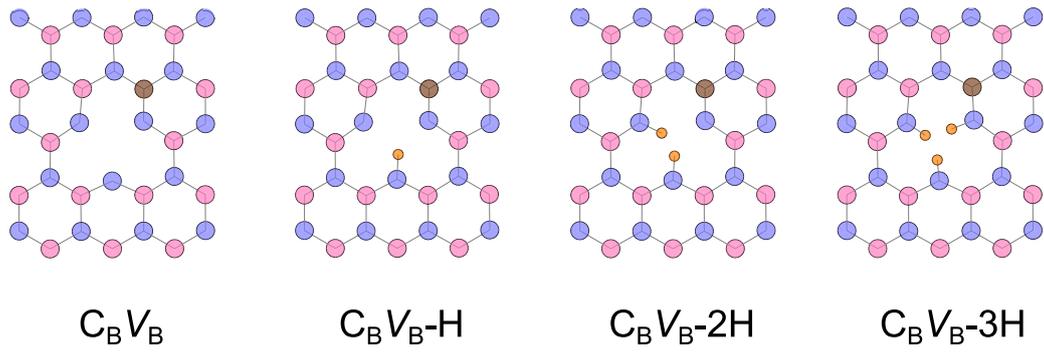

Fig. 3. $C_BV_B$-nH complexes in hBN. For $C_BV_B$-H and $C_BV_B$-2H, the most stable geometric configurations (i.e., those corresponding to the lowest energy) are shown. Blue spheres represent nitrogen atoms, pink – boron, brown – carbon, and orange – hydrogen.

a)

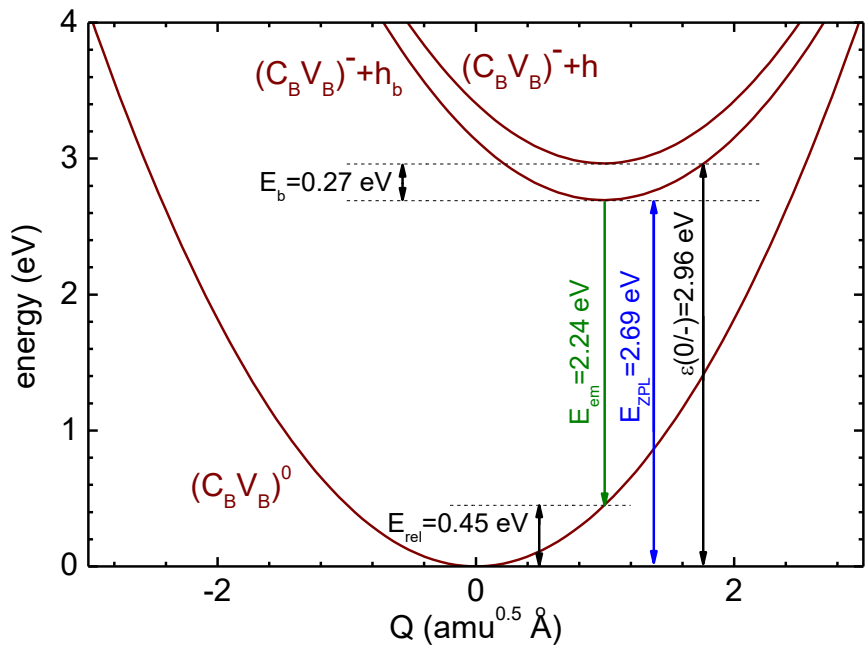

b)

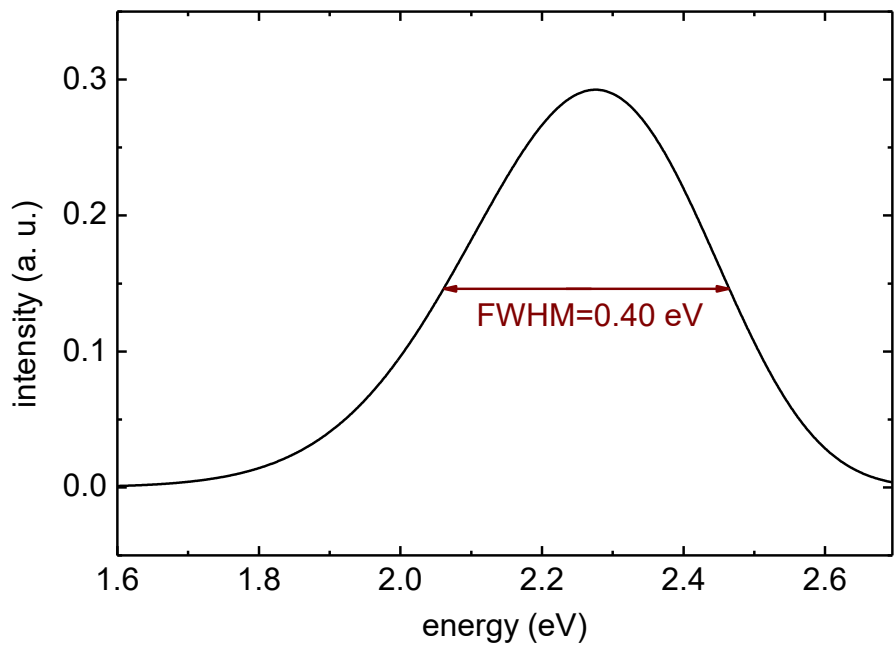

Fig. 4. (a) Configuration coordinate diagram describing optical emission following hole capture by negatively charged $C_BV_B$. (b) Calculated luminescence lineshape for hole capture by negatively charged $C_BV_B$, obtained using the single effective mode approach (with Gaussian smearing).